# THE 1979, MARCH 5 GAMMA-RAY TRANSIENT: WAS IT A CLASSIC GAMMA-RAY BURST?

E. E. Fenimore, R. W. Klebesadel, and J. G. Laros
Los Alamos National Laboratory, Los Alamos, New Mexico 87545, USA
e-mail: efenimore@lanl.gov

## ABSTRACT

The March 5th, 1979 gamma-ray transient has long been thought to be fundamentally different from the classic gamma-ray bursts (GRBs). It had recurrences, pulsations, and a soft spectral component unlike classic GRBs. With the exception of the soft component reported from the Konus experiment, the unusual characteristics of March 5th were detectable primarily because it was extremely bright. Computer limitations, satellite transmission effects or pulse pileup and dead-time effects have prevented, until now, the analysis of spectra from the International Comet Explorer (ICE) and the Pioneer Venus Orbiter (PVO). The ICE-PVO spectrum of the main peak differs markedly from the published Konus spectrum. Rather than being dominated by a soft component similar to that observed in the soft gamma repeaters (SGRs), the ICE-PVO spectrum appears to be consistent with a classic GRB spectrum, especially above 100 keV. Above 100 keV, the spectrum is consistent with thermal bremsstrahlung with a temperature of $\sim 200$ keV, somewhat soft but within the range of classic GRBs. We believe that, given the ICE-PVO spectral observations, the March 5th transient would have been classified as a classic GRB when it was discovered.

Although an formal analysis has not been done, the pulsations and recurrences might still be unique features that distinguish March 5th from GRBs. The ICE spectrum provides evidence for a low-energy cutoff at 100 keV. If high-velocity neutron stars are born as misaligned rotators with their velocities aligned with their spin axes and if their emissions are beamed, then when they are young their spatial distribution will be similar to the SGRs. If torques can align the field with the spin axis, then when they are old their spatial distribution will be isotropic like classic GRBs. Thus, the SGRs and GRBs could be consanguineous: high-velocity neutron stars initially produce SGR events (and, occasionally a GRB like March 5th) and when they are older and in the galactic corona, they go through a GRB phase. The March 5th event demonstrates that high-velocity neutron stars at distances of tens kpc are capable of producing events like classic GRBs.

This reanalysis has revised the March 5th intensity upward such that the peak luminosity at an assumed distance of 55 kpc is $1.9 \times 10^{45}$ erg s$^{-1}$. However, March 5th is consistent with the classic GRB log $N$-log $P$ distribution and is not necessarily extraordinarily bright.

*Subject Headings:* gamma rays: bursts - stars: individual (SGR 0526-66) - soft gamma repeaters - stars: neutron

# 1. INTRODUCTION

The 1979, March 5 transient (hereafter, "March 5th") was a pivotal event in gamma-ray astronomy (Mazets et al. 1979). It had characteristics which seemed to distinguish it from all other events seen before and since. An initial spike had a reported spectral component consistent with an optically thin thermal bremsstrahlung temperature of only $\sim 30$ keV (Mazets et al. 1979). This soft component contained half of the energy and 90% of the photons between 30 and 2000 keV. The initial spike was extremely bright: $\sim 1.5 \times 10^{-3}$ erg sec$^{-1}$ cm$^{-2}$, $10^4$ photon s$^{-1}$ cm$^{-2}$. A hard spectral tail contained an apparent redshifted annihilation line consistent with emission from the surface of a neutron star (Mazets et al. 1979). After the initial spike, the source pulsated with an 8 s periodicity for at least 200 s (Mazets et al. 1979, Barat et al. 1979, Terrell et al. 1980). The pulses were sinusoidal with a clear pulse/interpulse structure. These characteristics are strong evidence for a neutron star origin (Golenetskii et al. 1979).

The March 5th event was observed by nine satellites (Evans et al. 1980). The sharp rise of the initial spike provided an excellent fiducial resulting in an error box from triangulation which is still the most precisely resolved Gamma-ray burst (GRB) location (Cline et al. 1982). This location is consistent with a supernova remnant (N49) in the LMC (Evans et al. 1980). Assuming a distance of 55 kpc to the LMC, the peak intensity of the initial spike was $\sim 5 \times 10^{44}$ erg s$^{-1}$, a factor of $10^6$ times the Eddington luminosity for an unmagnetized neutron star.

In the days and years following the March 5th event, the source produced 15 more events, all soft and all $\sim 10^{-2}$ (or less) as intense as the March 5th event (Golenetskii, Ilyinskii, & Mazets 1984). Some of these events lasted up to several seconds and tended to be flat-topped.

Two other objects have shown similar behavior: soft spectra and recurrences. These objects have, therefore, been called the soft gamma repeaters (SGRs). The repeater known as SGR 1806-20 was discovered based on $\sim 110$ events (Laros et al. 1987, Atteia et al. 1987). Recently, this source became active again, and the Burst And Transient Source Experiment (BATSE) of the Compton Gamma-Ray Observatory (CGRO) detected six events (Kouveliotou et al. 1994), including one that was imaged by Asca. The Asca lo-cation was coincident with a supernova remnant, SNR G10.0-03 (Murakami et al. 1994), confirming an earlier suggestion (Kulkarni & Frail 1993). Another SGR (1900+14) produced three events in 1979 (Mazets, Golenetskii, and Guryan 1979) and was also later detected by BATSE (Kouveliotou et al. 1993). It, too, may be associated with an SNR (Hurley et al. 1994). These associations of SGRs with SNRs solidified the association of March 5th with the N49 SNR in the LMC. The identifications with SNR have provided estimates of the distance to these objects: SGR 1806-20 is at $\sim$ 15 kpc (Kulkarni & Frail 1993), and the March 5th source is at 55 kpc (Evans et al. 1980). When combined with their observed intensity, it is clear that these objects are extremely super-Eddington: the largest events from SGR 1806-20 had a total luminosity of $\sim 10^{42}$ erg s$^{-1}$ (Fenimore, Laros, & Ulmer 1994), a factor of $10^4$ above the Eddington luminosity for an unmagnetized neutron star. The March 5th event was even larger ($10^6$ times Eddington), although the recurrences from the March 5th source were comparable in intensity to the strongest events from SGR 1806-20. Recently, Rothschild, Kulkarni, & Lingenfelter (1994) reported a ROSAT source within the March 5th error box, whose position relative to the center of the SNR implies a *projected* velocity of 1200 km s$^{-1}$.

It has been argued that SGRs are another class of transients separate from x-ray bursts and gamma-ray bursts (Laros et al. 1987, Atteia et al. 1987). X-ray bursts are at lower energy ($\sim$ 3 keV *vs* 30 keV) and at much lower luminosity ($10^{38}$ *vs* $10^{42}$ erg s$^{-1}$). X-ray bursts display clear patterns in their recurrences (Lewin & Joss 1983) and are believed to be caused by thermonuclear flashes or accretion events, whereas SGRs seem to have no pattern in their recurrences (Laros et al. 1987). Gamma-ray bursts have a higher average photon energy (300 keV *vs* 30 keV) than the SGRs and do not obviously repeat. Both x-ray bursts and GRBs usually show spectral evolution as a function of time, whereas the SGRs do not (Kouveliotou, et al. 1987). Most important, the GRBs appear isotropic yet inhomogeneous, implying a distance much larger than typical galactic distance scales (Meegan et al. 1992). However, it is debatable whether GRBs come from cosmological distances or from an extended galactic corona of neutron stars (Lamb 1995, Paczyński 1995). The x-ray bursts and SGRs are probably associated with galactic disk distance scales (Laros et al. 1986, Kouveliotou et al. 1987). From these differences, it has been believed that SGRs are a class of events distinct from both x-ray bursts and gamma-ray bursts.

In this paper, we analyze the spectral observations made by the International Comet Explorer (ICE, formerly ISEE-3) and the Pioneer Venus Orbiter (PVO). We will argue that the March 5th event would have been classified as a classic gamma-ray burst when it was observed if the ICE-PVO analysis was available. This is a somewhat different issue than if it would still be classified as a gamma-ray burst (see section 5). All of March 5th's characteristics except its soft spectrum could have been the result of selection effects. Other single-spiked classic gamma-ray bursts could have had low-level pulsations but would be unobservable if they had a similar ratio of pulsations to peak intensity. Some multiple-spiked GRBs, like BATSE burst number 2151, could be bright enough that March 5th-like pulsations would be observable, although none have been. However, no formal analysis has been done on the BATSE events to determine if March 5th-like pulsations would be detectable in enough events to claim that the March 5th pulsations are unique. Similarly, the soft recurrences seen days to years later would not have been observable from other classic gamma-ray bursts if they had a similar ratio of intensity. Many classic gamma-ray bursts could have had pulsations, rapid rises, and recurrences, but only March 5th had sufficient strength to reveal them.

The soft spectrum was, indeed, a characteristic that classic GRBs do not exhibit. In Figure 1 we show as solid curves the diversity of gamma-ray burst spectra as observed by BATSE (cf. Band et al. 1993) normalized to each other at 100 keV. The dotted histogram is the March 5th spectral shape reported by Mazets et al. (1982) from the Konus experiment. The March 5 spectra is softer than any other classic gamma-ray bursts below 100 keV. The Band et al. (1993) spectra are averaged over most of the burst's duration so are probably somewhat softer than what one might obtain for the brightest peak. Also, short GRBs tend to have harder spectra. Thus, if one compared just short bursts or just the spectra of the brightest peaks in GRBs to March 5th, the disparity between the classic GRBs and the Konus March 5th spectrum would be even larger. The only classic gamma-ray bursts that come close to the March 5th spectrum are several events that have no change in slope. For example, GB910502 in Figure 1 is the spectrum closest to that of March 5th, but, it is effectively a power law with a single index. Most bursts have a change of slope such that they appear harder at low energies. GB910502 has a slope similar to the hard component in the March 5th Konus spectrum and only appears to be similar to the March 5th event because all of the events are normalized at 100 keV. Figure 1 demonstrates that the spectrum reported by Mazets et al. (1982) for March 5th is indeed inconsistent with the spectra of the classic gamma-ray bursts. Also shown in Figure 1 as a dotted curve is the spectrum of SGR 1806-20 (from Fenimore, Laros, & Ulmer, 1994). The early work on SGR 1806-20 reported (Atteia et al. 1987) an effective temperature of 35 keV, the same as the soft component of the initial peak of March 5th. Subsequent refinements of the analyses for SGR 1806-20 (Fenimore, Laros, & Ulmer, 1994) revealed a somewhat softer spectrum: 22 keV. Thus, although the histogram and dotted curve are different in Figure 1, it has long been believed that March 5th had a soft component very similar to the SGRs and unlike the classic GRBs.

The Konus detector covered a low energy range (down to 30 keV) and had a relatively large area (50 cm$^2$). It had a fairly direct view of the March 5th event. It is unclear what precautions were taken to accommodate dead-time and pulse pileup effects. Although the Konus experiment had shaping amplifiers with very short time constants (1 microsecond), the conversion to pulse height had a rather long dead time of 1 ms (Mazets, private communication). This implies large dead-time effects (although much smaller pulse pileup effects) since the counting rate during the event was $2 \times 10^5$ cts s$^{-1}$ (Mazets et al. 1982). The Konus spectral sample was taken over a 4 s period whereas the peak lasted only 0.2 s. Since the peak was $\sim$ 100 times brighter than the subsequent pulsations, the 4 s period would be dominated by the counts from the peak *if the dead-time is short.* Assuming a 1 ms dead-time, the electronics limited the photons from the peak to $\sim$ 200, whereas the pulsations contributed $\sim$ 4000 photons. Thus, the long dead time emphasizes the soft contribution of the pulsations and leads to a spectrum with a soft component. It appears that if our ICE spectrum for the initial peak plus 4 s of the pulsations are folded through the Konus electronics, we would predict a spectrum much like that reported by Mazets et al. (1982): 10% of the photons in a hard distribution (from the initial peak) and 90% of the photons with a soft distribution (from the pulsations).

Although nine instruments detected March 5th, most have not reported a spectrum because of uncertain dead-time and pulse pileup effects, or because the event passed through the body of the satellite, introducing uncertainties in the response function. Only Konus and ICE observed the initial peak below 100 keV. The PVO instrument was sensitive above 100 keV. The PVO and ICE spectral data complement each other. Because PVO viewed the event directly, it had dead-time and pulse pileup effects but little uncertainty in the detector response function. ICE viewed the event through the spacecraft, so had small dead time and pulse pileup, but the detector response function is somewhat uncertain. We have combined the two data sets to obtain a consistent spectrum.

## 2. ICE OBSERVATIONS OF MARCH 5TH

ICE had a NaI detector collimated at low energies to observe close to the ecliptic plane (see Anderson et al. 1978). Because the detector was located on the opposite side of the satellite from the March 5th location, most of the low-energy photons were blocked such that ICE did not have strong dead-time effects. The March 5th event occurred within 42 arcminutes of the nominal rotation axis of the satellite, at the south ecliptic pole. (This has about the same *a priori* probability as the March 5th error box containing an SNR.) Fortunately, the center of the ICE satellite was a nearly empty, hollow tube (T. von Rosenvinge, private communication). The detector was centered on this tube and had a diameter much smaller than the tube such that the walls of the tube were far from the line of sight. Thus, we model the signal seen by the detector as two components: a direct signal that comes up the tube and a scattered signal from the rest of the satellite. We treat these two components separately in our Monte Carlo calculation of the response matrix. The detector assembly was well modelled from precise blueprints.

### 2.1 ICE Response Matrix

The central tube was empty except for structural plates and a small accelerometer in the line of sight, as well as a few boxes out of the line of sight. However, to be conservative, we have verified that our results do not depend on the exact amount of material in the line of sight. For example, it is possible that a cable bundle or some late addition to the hardware was placed in the line of sight, since, unlike the collimated field of view, the material within the tube was not as well controlled or documented. Our best estimate of the amount of material in the line of sight is equivalent to 1.0 cm of aluminum. To accommodate the possibility that our estimate is wrong, we have generated response matrices with a range of thicknesses that encompasses all reasonable amounts of materials that might be in the tube. We used five different column densities, corresponding to 1/4, 1/2, 1, 2, and 4 cm of aluminum. We have also investigated much thicker and thinner amounts of material and none are consistent with the Konus observations. These column densities do not include the material associated with the gamma-ray spectrometer such as the electronics and the photomultipler tube, which were modelled separately. We have also experimented with clumps of materials within the tube but find that the response matrix is only sensitive to the total amount of material in the tube.

The other contributor to the response matrix is the mass of the satellite acting as a scatterer. The gamma-ray spectrometer was positioned on a tower such that most of the mass of the satellite was more than a meter away. Such a distance greatly decreases the impact of the satellite on our response matrices. We have modelled the major structural elements, the solar panels, and we have filled the equipment bays with solid material with a density such that the mass is equivalent to the actual mass. We have found that, as far as the detector is concerned, the response is sensitive only to the total mass of the satellite. We have used 400 kg for the mass, not including the central tube or the gamma-ray spectrometer (T. von Rosenvinge, private communication). We have also generated response matrices for the extreme cases of 250 kg and 500 kg to demonstrate that the exact mass of the satellite body is not important. Neither allows a spectral shape with an SGR-like soft component. We also experimented with an equipment bay consisting of constantly changing, randomly placed, solid lumps (rather than diffuse material with an equivalent mass) such that photons could sometimes interact closer to the detector before their first scattering. We wanted to verify that empty paths through most of the satellite would not change the effective area of the detector. Indeed, all mass distributions gave the same response matrix if they had the same total mass.

The exact chemical makeup of the material within the satellite (or the tube) is not important. Different material would produce different fluorescent photons and have absorption edges at different energies. However, the fluorescent photons and edges are at low energy such that, for most materials, those photons are absorbed before reaching the detector. Photoelectric absorption, Compton scattering, and pair production for low atomic number material are self-similar such that varying the overall column density accommodates variations in chemical makeup. For simplicity we assume that all of the satellite material is aluminum.

### 2.2 Trial Spectrum Method

During the March 5th event, the ICE detector collected data in two ways. The sudden increase of counts triggered a special memory to record high-resolution temporal data. The triggered data covered 322 keV to 3190 keV with 6 channels and 0.05 sec time resolution. We do not use these data because of the lack of coverage at low energy. Also available for significant time before and after the event are data covering 26 to 3190 keV with 12 channels and 0.5 sec time resolution for most channels. We did not use the lowest energy channel because its counts were due only to Compton scattered photons from the full energy range; it did not have a photoelectric peak in its response matrix. Thus, we used 11 channels covering the range of 43 to 3190 keV. The main peak of March 5th lasted $\sim 0.2$ sec and was entirely contained within one of the 0.5 sec samples. The overall count rate was 20630 Hz. At such a count rate, the average dead-time correction (including the effects of the time variations in the peak) was 7.5%. The background was 530 Hz, which had little affect on the signal. Later, we will compare the ICE data to the spectrum reported by Mazets et al. (1982), which was recorded over a 4 sec period. Since the ICE peak signal dominates the background (and the subsequent pulsations), the difference in recording time affects only the overall normalization of the spectrum, an effect for which we account.

We use a "goodness of fit" criterion to determine the best-fit parameters of a trial spectrum. Let $O_i$ be the gross observed counts in channel $i$, and $B_i$ is the expected number of background counts during the sample. We define

$$\chi_j^2 = \sum_i \chi_{ij}^2 = \sum_i \frac{[O_i - B_i - \int R_{ij}(E)\phi(E)dE]^2}{\sigma_{O_i}^2 + \sigma_{B_i}^2 + \beta^2 O_i^2} \quad , \qquad (1)$$

where $R_{ij}(E)$ is the effective area of the $i$-th channel for the $j$-th response matrix (e.g., $j$ denotes using 1/4, 1/2, 1, 2, or 4 cm of aluminum in the central tube). The trial photon spectrum is equal to $\phi(E)$; $\sigma_{O_i}^2$ is equal to $O_i$; and $\sigma_{B_i}^2$ is based on the background, but it is small because the background is determined over a long period of time (e.g., 10 sec). The intent of the $\beta^2 O_i^2$ term is to account for systematic uncertainties in the response matrix. March 5th was very bright, and if one blindly assumes Poisson statistics, some channels would supposedly be known to a few percent. In the case of ICE, where we have an uncertain geometry, we do not know the response matrix better than a few percent. Without $\beta$, the fits gravitate to the apparently best determined points although those points are as uncertain as the rest of the points due to the systematic uncertainties. The $\beta$ term limits the signal-to-noise ratio for a channel to a minimum of $\beta^{-1}$. Unless otherwise stated, we will use $\beta = 0.05$; that is, we assume the systematic noise is 5%. We selected this value for two reasons: First, we estimate that we probably have 5% systematic uncertainties in our response matrices and normalizations. Second, when 5% is used, the best-fit $\chi^2$ values usually have a value the order of the number of degrees of freedom. Although $\beta$ prevents unrealistic estimates of the uncertainty associated with the measurements, the resulting "$\chi^2$" can be used only relatively; neither the probability of $\chi^2$ occurring by chance nor the confidence regions can be properly estimated. However, it also is not possible to calculate the chance probability or confidence region if $\beta$ is not used since the systematic uncertainties are not accounted for.

### 2.3 Konus Folded with ICE

Figure 2 plots the fits of reported spectrum of March 5 main peak (cf. Mazets et al. 1982) to our ICE observations. We used the following trial spectrum (in photon sec$^{-1}$ cm$^{-2}$ keV$^{-1}$):

$$\phi(E) = 3.03 \times 10^4 E^{-1} e^{-E/35} \qquad (2)$$

$$+ 10^3 E^{-1} e^{E/520}$$

$$+1.028 e^{\left(\frac{E-430}{48}\right)^2} \quad ,$$

where E is in keV. The first two terms are provided by Mazets et al. (1982, Fig. 2). We determined the third term (the emission line at 430 keV) from fits to Figure 2 of Mazets et al. (1982). In our fitting of equation (2) to the ICE observations, we allowed one free parameter: an overall scale factor. The bottom panel gives the residues ($\chi_{ij}$), that is, the difference between the observations ($O_i - B_i$) and the model in units of the uncertainties. By definition, the size of the error bars on the points in the bottom panel is $\pm 1$. In the bottom panel we show the residuals assuming five different response matrices (corresponding to 1/4, 1/2, 1, 2, and 4 cm of aluminum in the tube). None provide an acceptable fit, $\chi^2$ ranges from 370 to 510. Rather than residuals that scatter about zero, there is a strong trend demonstrating that equation (2) is a shape that is inconsistent with the ICE observations. This inconsistency is not due to our uncertainty in the ICE response matrix, since the matrices that we are using span effectively all possibilities. Although the inconsistency is strongest near $\sim 300$ keV, it is not due to the emission line at 430 keV, which only contributes $\sim 1\%$ to $\chi^2$.

### 2.4 Best Spectral Fit from ICE

To investigate what spectral shapes are consistent with the ICE data, we have used a trial spectrum built from three connected power laws:

$$\phi(E) = a_1 (E/E_1)^{\alpha_1} \qquad E < E_1 \qquad (3a)$$

$$= e^{m \ln E + b} \qquad E_1 < E < E_2 \qquad (3b)$$

$$= a_2 (E/E_2)^{\alpha_2} \qquad E_2 < E \quad , \qquad (3c)$$

where

$$m = \frac{\ln(a_2) - \ln(a_1)}{\ln(E_2) - \ln(E_1)} \qquad (3d)$$

and

$$b = \ln(a_1) - m \ln(E_1) \quad . \qquad (3e)$$

This functional form was selected because it is flexible enough to make many different shapes. In the fitting process, we initialized the search for the $\chi^2$ minimum with a shape that is similar to the Konus shape, that is, a spectrum that is concave upward ($\alpha_1 < \alpha_2$). No $\chi^2$ minimum or even a secondary minimum existed with a such a shape. The best-fit shape is concave downward ($\alpha_1 > \alpha_2$) above 100 keV and has very little emission below 100 keV. In the top panel of Figure 3, the solid curve is the best-fit spectral shape for the response matrix using 1 cm of aluminum in the tube and 6 free parameters ($a_1$, $\alpha_1$, $a_2$, $\alpha_2$, $E_1$, and $E_2$). The solid squares in the bottom panel give the residuals for the best fit, the resulting $\chi^2$ is 23.2. The concave downward spectrum fits much better than the concave upward spectrum of Figure 2.

Often, a variety of spectral shapes can all be consistent with the observations, yet they can be quite different. This "obliging" nature of trial spectrum fitting is intrinsic to any detector that does not have a purely diagonal response matrix (see Fenimore, Klebesadel, & Laros 1983). We investigated extreme cases that might be consistent with the Konus spectrum. We found the lower energy slope ($\alpha_1$,) which caused $\chi^2$ to increase by 7.0, 12.75, and 20.0 whereas the other 5 parameters were allowed to vary. These spectra lie on the $1\sigma, 2\sigma$, and $3\sigma$ confidence surfaces and are plotted in Figure 3 as the dotted, dashed, and long dashed curves, respectively. The spectrum above 100 keV is well within the range of classic GRBs spectra. Although the best-fit spectrum shows a steep cutoff below 100 keV (unlike typical GRB spectra), the confidence region for the acceptable shapes include spectra very much like GRB spectra. Figures 1 and 3 are at the same scale and overlaying them shows that the $2\sigma$ ICE-PVO spectrum falls within the range of observed GRB spectra. However, we do not find even extreme fits that are consistent with the soft SGR-like spectral component reported for March 5th from the Konus experiment.

A $\Delta\chi^2$ value of 7.0 is very conservative; the parameter space that has $\Delta\chi^2 < 7$ contains the true model in at least 68% of occurrences. It is conservative because it assumes no coupling between the parameters, whereas we have strong coupling (e.g., changing $\alpha_1$ changes $a_1$). When the $\chi^2$ surface corresponding to a $\Delta\chi^2$ value is projected onto an axis associated with a single parameter, the range of the parameter is larger when parameters are coupled so the range contains more than 68% of the volume (see Lampton, Margon, & Bowyer 1976). To demonstrate that it is conservative, we simulated 128 realizations of the best-fit spectrum (i.e., the solid curve in Fig. 3) and analyzed them. All resulting best-fit spectra were enclosed within the envelope of the dotted line in Figure 3. We conclude that neither the obliging nature of the analysis nor the statistical variations can modify the best-fit ICE spectrum enough to make it consistent with the Konus spectrum.

Next, we investigate whether uncertainties in the response matrix could be responsible for the discrepancy between Konus and ICE. Figure 4 shows the best-fit ICE spectrum, assuming a range of response matrices. From top to bottom, the curves correspond to 1/4, 1/2, 1, 2, and 4 cm of aluminum assumed to be in the central tube. For each response matrix, we have also found the $1\sigma$ limit as was done in Figure 3. These ranges appear similar to that shown in Figure 3. In the bottom panel of Figure 4 we show the residues. The corresponding $\chi^2$ values range from 23.2 to 24.2, substantially better than that found when we used the Konus spectrum (370 to 510). We conclude that systematic errors in the response matrix cannot modify the best-fit ICE spectrum enough to make it consistent with the reported Konus spectrum.

From Figures 1, 2, and 3, it is clear that the ICE March 5th spectrum shares more similarities with the spectra of classic GRBs than with the soft gamma repeaters. Although most classic GRBs do not show roll-overs below 100 keV, such roll-overs have been reported for several classic GRBs: GB781119 (Barat 1983), GB820328b (Mazets et al. 1983), GB910709 (Band et al. 1993 and Fig. 1), and GB911007 (Pendleton et al. 1993). Thus, whereas the Konus March 5th spectral shape falls outside the range of GRBs, the ICE shape is within the range of the classic GRB shapes, especially above 100 keV.

### 3. PVO OBSERVATIONS OF MARCH 5TH

PVO had two small CsI detectors (22 cm$^2$ total effective area, see Klebesadel et al. 1980) directly facing the south ecliptic pole near where March 5th originated. As such, there is effectively no uncertainty in how the scintillator responds to an incident spectrum. PVO has four energy channels: 100-200, 200-500, 500-1000, and 1000-2000 keV. When the background is nearly constant, counts are reported with a time resolution that depends on the satellite telemetry mode, often about once per second. A statistical significant variation (11.2 $\sigma$) in the background count rate causes a special memory to record the burst with high time resolution. In this "trigger" mode, counts in the full energy range (100-2000 keV) are reported every 11.72 msec and counts in the four energy channels are reported every 187.5 msec. For sufficiently high count rates, there is a "time-to-spill" mode where the time to record 16 counts is reported rather than the number of counts within 11.72 msec (see Klebesadel et al. 1980 for more instrumental details). The March 5th event was extremely bright, such that the time-to-spill mode provided temporal resolution of ∼0.5 msec and spectral resolution with ∼ 6 msec resolution during the main peak.

### 3.1 PVO Dead-time and Pulse Pileup Effects

PVO had an unusual dead-time and pulse pileup situation. The four spectral channels of PVO were defined by level discriminators, but the logic gave priority to higher energy photons. This causes a coupling between pulse pileup and dead time. Photons can occur so frequently that several produce simultaneous electronic pulses within the pulse shaping. The heights of these pulses accumulate, mimicking a higher energy photon. This is a common feature of pulse-processing logic at high counting rates. Furthermore, it takes a finite time to process the pulses and, while it is processing one pulse (or a sum of several), other pulses are not counted. The result is "dead time", that is, time when the instrument misses photons. Since the PVO higher-energy channels have priority over lower-energy channels, there is an inverse correlation between energy and dead time. This is unlike most pulse-processing logic, where all energy channels have the same dead time.

To accommodate the coupling between dead time and pulse pileup, we analyze spectra by folding trial spectral shapes through the instrumental response and the electronics until a $\chi^2$ statistic is minimized. To incorporate the PVO dead time and pulse pileup, we generate each candidate spectrum by modelling a large number of photons one at a time through the scintillator response and the electronics. We select each photon's energy and time of occurrence from Poisson distributions. The photon rate was varied to match the observed temporal structure in the peak. The electronic pulse from each photon is added to other pulses, if any, that co-exist in the circuitry. The summed electronic signal is converted to events in the four PVO energy channels just like the flight electronics. It takes ∼ 80,000 simulated photons to generate a candidate spectrum, 10 times more photons than was observed. The Monte Carlo statistics, therefore, do not substantially affect the $\chi^2$ statistic. In a typical resulting spectrum, the highest-energy channel has about 35% more counts than actually occurred in the spectrum as a result of pulse pileup (with little dead time) and the lowest-energy channel (100 to 200 keV) has about 30% fewer photons than actually occurred. (These percentages are averaged over the 0.2 s duration; they are higher during the brightest portion of the peak, an effect for which we account.) In the case of the lowest-energy channel, pulse pileup puts more counts into the channel, but dead time reduced the number that can be counted. The result is a net reduction in the number of counted events. The PVO spectrum basically agrees with the ICE spectrum. This, plus the non-excessive dead time in PVO, indicates that we have sucessfully accounted for all electronics effects. We feel that we can accurately model the dead time although it requires a substantial amount of computer time to do so. (We have used more than $2 \times 10^7$ sec of time on our modern machines, so even if this work were started on 1979-class machines, the analysis would not have been completed until recently.)

*3.2 Best Spectral Fit from PVO*

Figure 5 shows the fit of a power law spectrum to the PVO March 5th observations of the first 200 ms of the initial peak. The dotted curve in the top panel and the open squares in the bottom panel give the best-fit power law. The fit is very poor; $\chi^2$ is 155. This is unusual; a single power law can usually give a better fit to the four PVO channels. The solid line and solid squares in Figure 5 are for a single power law with a high-energy cutoff and gives a $\chi^2$ of 10.0. Since the inclusion of one new parameter reduces the $\chi^2$ from 155 to 10, the additional parameter is certainly justified even though $\beta$ is not zero in equation (1). Less steep cutoffs fit better. We have fit two connected powers to the PVO data:

$$\phi(E) = a_1 (E/E_c)^{\alpha_1} \qquad\qquad E < E_c \qquad\qquad (4a)$$

$$a_1 (E/E_c)^{\alpha_2} \qquad\qquad E > E_c \quad . \qquad\qquad (4b)$$

There are four free parameters (one more than the solid line in Fig. 5): the connecting energy ($E_c$), the value of the spectrum at the connecting energy ($a_1$), and the slopes above and below the connecting energy ($\alpha_1, \alpha_2$). The best fit has a $\chi^2$ of 0.03, as expected, since there are only four observations and no degrees of freedom (see the dashed line in the top panel and the triangles in the bottom panel of Fig. 5). A thermal bremsstrahlung shape ($E^{-1}e^{-E/kT}$) fits a little better than the power law with a cutoff: $\chi^2$ is 6.5 for $kT = 246$ keV (see the long dashed curve and open circles in Fig. 5).

## 4. COMPARISONS OF ICE, PVO, AND KONUS

Figure 6 shows a joint fit of ICE and PVO. The trial spectrum was a power law connected to the thermal bremsstrahlung shape. We selected this shape because thermal bremsstrahlung often fits GRB spectra (although it is not believed to be the operative physical process) and because it is one of the shapes that successfully fit the SGR 1806-20 spectra (see Fenimore, Laros, & Ulmer 1994). We used

$$\phi(E) = \frac{ae^{-E_c/kT}}{E_c}\left(\frac{E}{E_c}\right)^\alpha \qquad E < E_c \qquad (5a)$$

$$\phi(E) = \frac{ae^{-E/kT}}{E} \qquad E > E_c \ . \qquad (5b)$$

Here, we have characterized the thermal bremsstrahlung shape by a "temperature" ($kT$), a convenient parameterization that allows comparisons to previous work. For Figure 6, we used an ICE response matrix corresponding to 1 cm of aluminum in the central core; the solid curve is the best-fit spectrum. We used $\beta = 0.10$ since there is the additional systematic effect of comparing two different instruments. In the bottom panel, the open squares are for PVO, and the solid squares are for ICE. The resulting $\chi^2$ is 17.4. The best-fit $kT$ is 202 keV, which is lower than a typical GRB but well within the range of "temperatures" seen for GRBs. The agreement between ICE and PVO in Figure 6 is as good as we usually obtain for GRBs seen by these two instruments. We find the $1\sigma, 2\sigma$, and $3\sigma$ confidence values for the low-energy slope as we did in Figure 3. They are plotted as dotted, dashed, and long-dashed curves in Figure 6.

Equation (5) was also used to fit to the spectrum of SGR 1806-20 with the results that $E_c = 14$ keV and $kT = 22$ keV (Fenimore, Laros, & Ulmer 1994). The March 5th spectrum requires $E_c = 155$ keV and $kT = 202$ keV.

Thus, the March 5th spectrum is very similar to the SGR 1806-20 spectrum that has been shifted by a factor of $\sim 10$. The higher luminosity of March 5th could have caused relativistic bulk motion which shifted the spectrum.

We have also repeated the analysis shown in Figure 6 for the other ICE response matrices. Figure 4 shows that the overall normalization of the spectrum increases with an increasing amount of material in the central tube of ICE. The PVO observations have no such flexibility. Effectively, the PVO spectrum verifies that the correct ICE response matrix is being used. The range of response matrices give $\chi^2 = 28.0, 24.2, 17.4, 21.0$, and $98.0$ for 1/4, 1/2, 1, 2, and 4 cm of aluminum, respectively. Thus, our choice of 1 cm of aluminum in the central core also gives the lowest $\chi^2$ when both ICE and PVO are used. It would be possible to treat the density of the material in the central tube as a free parameter and reduce the $\chi^2$ for the PVO-ICE fit, but each calculation requires $\sim 10^6$ seconds of computer time.

The peak intensity of the March 5th event is of interest since it exceeds the Eddington limit by many orders of magnitude. Table 1 summarizes the peak intensities (photons cm$^{-2}$ s$^{-1}$) and corresponding luminosities. ($L_{45} = L_0 10^{-45}$ erg s$^{-1}$), assuming a distance of 55 kpc and that it radiates into $4\pi$. (Although recent work has indicated that the LMC is at $\sim 50$ kpc, we will use 55 kpc to facilitate comparisons to the earlier work on March 5th.) Various reports have used different energy ranges and time samples. For example, Konus observations are for the 30 to 2000 keV range and averaged over 0.2 s. In Table 1, we use three energy ranges: the 30 to 2000 keV range corresponds to Konus; the 50 to 300 keV range corresponds to BATSE; and the 100 to 500 keV range was used in the combined BATSE-PVO log $N$-log $P$ distribution. The excellent time resolution of PVO allows us to find the peak intensity on 10 ms and 64 ms time scales. The 10 ms sample is interesting because it is much longer than the dynamical time scale of a neutron star so the corresponding luminosity (e.g., $1.9 \times 10^{45}$ erg s$^{-1}$ in 30 to 2000 keV) is the amount which models must explain in the context of an Eddington limit. Note that it is a factor of 4 larger than the often-quoted Konus luminosity (Mazets et al. 1982), which was based on an average over 200 ms (i.e., $5 \times 10^{44}$ erg s$^{-1}$). How difficult it is to accommodate a super Eddington flux is not clear; several recent papers have suggested methods in the context of the SGRs (Paczyński 1992, Ulmer 1994, Thompson & Duncan 1995, Miller 1995). The 64 and 256 ms samples are useful because they are the sample periods reported by BATSE. Since most of the emission was in the initial peak, the intensities for the 1024 ms sample period of BATSE are $\sim 1/4$ the value found for 256 ms. Of special interest is the peak intensity in the 100 to 500 keV range sampled over 256 ms because this is the appropriate value that would be used in the combined BATSE-PVO log $N$-log $P$ distribution (see Fig. 1a in Fenimore et al. 1993). Even though we determine March 5th to be brighter than reported from Konus, it is consistent with the GRB log $N$-log $P$ distribution. In the combined BATSE-PVO log $N$-log $P$ distribution, the brightest event (GRB900808) is 900 photons sec$^{-1}$ cm$^{-2}$, whereas the appropriate ICE-PVO value for March 5th is 3100 photons s$^{-1}$ cm$^{-2}$ (see Table 1). We have simulated the combined BATSE-PVO log $N$-log $P$ and found that GRBs as bright as or brighter than March 5th have a 12% chance of occurring within the life of PVO. In fact, March 5th is not necessarily the brightest event ever seen. The GRB on 1983 August 1 (Boer et al. 1992) was clearly a classic burst: the duration was $\sim 50$ s and the spectral power ($E^2\phi[E]$) peaked at $> 500$ keV. (PVO did not observe this event.) There was an initial bright peak that lasted 3 s and whose spectrum (i.e., Fig. 3 in Boer et al. 1992) exceeded the ICE-PVO spectrum for March 5th over most of the energy range of 30 to 2000 keV. Although no intensity was quoted for the event, it appears to be brighter than March 5th. Thus, the March 5th event is fully consistent with the classic GRB log $N$-log $P$ distribution and is not necessarily extraordinarily bright.

Of course, March 5th could, in fact, be extraordinarily bright and just happen to have an intensity consistent with the log $N$-log $P$ distribution. For example, consistency with the log $N$-log $P$ distribution does not take into account that the March 5th distance implies a larger luminosity than expected from some models of a galactic corona of neutron stars.

We agree with Mazets et al. (1979) that the pulsations following the initial peak were substantially softer, the photon number index for the pulsations in PVO is $-3.9$. We note that the pulsations are weak enough that no instrument had pulse pileup or dead-time effects. We can neither confirm nor refute the existence of an emission line near 430 keV. Including such a line in the best-fit function for the ICE fit increases $\chi^2$ from 23 to 27 (with no change in the number of degrees of freedom). Since our $\chi^2$ values are relative and the change is small, we cannot reject the presence of a line. However, we feel that one must have a reliable continuum before the reality of a line can be established. We note that, when there is dead time, the statistics on the samples in the spectrum should be based on the number of recorded events rather than the number of dead-time corrected counts. The error bars in Figure 2 of Mazets et al. (1982) appear to be based on the dead-time corrected number of counts. If the error bars are, in fact, larger, the redshifted annihilation line might be a statistical fluctuation.

Besides the spectra from Konus (Mazets et al. 1979, 1982), only two other instruments have reported spectral information on March 5th. We reported hardness ratios from PVO (Fenimore et al. 1981) in which we attempted an analytic deadtime correction. The Monte Carlo approach used in this paper is more accurate. At the time of the Fenimore et al. 1981 paper, we thought the PVO electronic gain was lower, so the calibration between the PVO hardness ratio and equivalent blackbody was incorrect. This led us to incorrectly conclude that March 5th was consistent with a characteristic temperature of $\sim$ 60 keV. The only other report of a spectrum from March 5th occurred as a private communication from K. Hurley in Norris et al. 1991. However, that spectrum (from one the Venera SIGNE experiments), has not been corrected for detector efficiency, pulse pileup, or deadtime and should not be considered reliable (private communication, K. Hurley). The spectrum did not go below 250 keV, so it cannot address the issue of the presence of a soft SGR-like component.

## 5. DISCUSSION
### 5.1 March 5th: a New View

Originally, March 5th was considered the prototype for the soft gamma repeaters since it was the first event to have a strong 30 keV spectral component and recurrences. Norris et al. (1991) argued that the initial peak of March 5th was so much more luminous than the March 5th recurrences and other SGRs (such as SGR 1806-20) that the initial peak should not be considered an SGR event. In this paper, we have shown that the initial peak did not have a soft SGR-like component and that the spectrum is within the range of classic GRBs, although somewhat soft. In Fenimore (1995), we investigated the variations in hardness within the peak of the March 5th event and found that, at times, the event is twice as hard as the spectrum reported here for the average over the peak. We believe that, if it had been possible to do this analysis when March 5th was discovered, the event would have been classified as a classic GRB. The other so-called unique characteristics (sharp rise, pulsations, recurrences) would all have been attributed to selection effects since many other GRBs could have had these characteristics, but they would be undetectable in bursts substantially weaker than March 5th.

The brightness of March 5th is extreme, but, perhaps, not extraordinary. Originally, it was believed that GRBs were at distances the order of 100 pc such that, if March 5th was a GRB at 55 kpc, it would be $10^6$ times too bright. If GRBs are cosmological and March 5th is a GRB, then it is $10^6$ times too dim. The March 5th luminosity is closer to that expected for galactic corona models. The appropriate March 5th luminosity to compare to the galactic corona models which have been fit to the BATSE data is based on the 50 to 300 keV bandpass and 256 ms sample period (i.e., $2.7 \times 10^{44}$ erg s$^{-1}$, Table 1). Such models require luminosities of a few times $10^{42}$ erg s$^{-1}$ (D. Lamb, private communication, Li & Dermer 1992) Thus, if in the galactic corona, GRB luminosities are about within a factor of 100 of March 5th. Based on the curvature of the BATSE-PVO log $N$-log $P$ distribution, the GRB luminosity function can have a range of $\sim 10$ (Ulmer, Wijers, & Fenimore 1995). All of the log $N$-log $P$ studies and the galactic corona models assume a standard candle luminosity. If fluence is, in fact, standard instead of luminosity, the short events (such as March 5th) would appear proportionally brighter than under the standard luminosity assumption. Based on fluence, March 5th is the 13th brightest burst in PVO, a factor of 6 weaker than the brightest burst. Thus, there are perhaps several ways for the March 5th event to be consistent with both the classic GRB log $N$-log $P$ distribution and the luminosity required in galactic corona models.

It has been claimed that "unique" features of March 5th distinguish it from classic GRBs (Cline et al. 1980): (1) extreme intensity, (2) rapid rise, (3) soft spectrum, (4) soft tail, (5) pulsations, (6) repetitions, and (7) an object in its error box. In this paper we have shown that the spectrum was, in fact, as hard as some classic GRBs. Its intensity is no longer extraordinary because the passage of time has not given us anymore such events. In Fenimore (1995), we show that the March 5th rise time is $\leq 1$ ms rather than the previously reported 0.2 ms (cf. Cline et al. 1980). Other classic GRBs have similar rise times. Fenimore (1995) also points out that the hardness ratio variations within the peak of March 5th are similar to that seen in classic GRBs. Soft tails have been found in many GRBs (Murakami et al. 1992). Although there have been some hints of repetitions (Wang & Lingenfelter 1993, Quashnock & Lamb 1993) in classic GRBs, the repetitions, pulsations, and an object in its error box are the only features of March 5th that remain unique. Below, we speculate why March 5th might show pulsations and an object within its error box whereas, GRBs do not.

### 5.2 Is March 5th a GRB?

It is crucially important to determine whether the March 5th event should be considered a classic GRB. If March 5th is a classic GRB, then we have partially answered the question as to whether GRBs are cosmological or galactic since the recurrent events of March 5th are clearly associated with the SGRs that have a distance scale of tens of kpc. Due to their age and positions relative to supernova remnants, the SGRs seem to be associated with high-velocity neutron stars (Duncan & Thompson 1992, Kulkarni & Frail, 1993, Hurley et al. 1994, Rothschild, Kulkarni, & Lingenfelter 1994). In fact, March 5th is the best candidate for a high-velocity neutron star associated with an SGR (cf. Rothschild, Kulkarni, & Lingenfelter 1994). High-velocity neutron stars are also required to produce an extended galactic corona that is consistent with the BATSE isotropy (Li & Dermer 1992, Podsiadlowski, Rees, & Ruderman 1995). Recently, pulsar velocities have been revised upwards such that it is clear a population of neutron stars exists that can make the galactic corona of GRB sources (Lyne & Lorimer 1994). If March 5th is a GRB, then the SGRs and GRBs are consanguineous: high-velocity neutron stars initially produce SGR events (and occasionally a GRB like March 5th), and when they are older and in the galactic corona, they go through a GRB phase. These neutron stars could differ from those that produce radio pulsars and x-ray bursts perhaps because they have a much higher magnetic field, $10^{15}$ Gauss, as implied by the spin period of March 5th (Duncan &

Thompson 1992).

One of the deepest mysteries in high-energy astrophysics is the distance scale to the classic GRBs (Lamb 1995, Paczyński 1995). If one concludes that March 5th is a classic GRB, then that is equivalent to concluding that at leaset some GRBs are galactic. Thus, the most important question concerning March 5th is whether it was a galactic GRB. There are two directions one can take: either the unique features imply that March 5th is not a GRB, or the similarities imply that it is.

*Do the unique features of March 5th distinguish it from GRBs?* GRBs do not appear to have the pulsations or repetitions associated with March 5th. There are three reasons why the differences might not disqualify March 5th as a galactic GRB. First, there has been no analysis of the BATSE events examining whether March 5th is inconsistent with them. Thus, it is not known whether these differences would be detectable in bursts substantially less intense than March 5th. One must artificially increase the noise in March 5th until it is commensurate with the short BATSE events and then evaluate whether one can, indeed, distinguish March 5th from the events seen by BATSE. Although such work is in progress, at this point one cannot be sure that the March 5th characteristics are unique. Second, even if the pulsations and repetitions are unique, they could represent a separate type of activity that was initiated by the GRB (i.e., the initial peak). Neutron stars might have both SGR and GRB phenomena. Repetitions and pulsations could be a common feature of the SGR phase but not the GRB phase. In the case of the March 5th event, a GRB event occurred, and it was followed by an SGR event. Third, the March 5th event, in the context of galactic corona models for GRBs, is $\sim 10^2$ times brighter than the typical GRB. This larger luminosity could have caused the unique characteristics of the initial peak of March 5th: the sharp rise and subsequent pulsations.

*Do the similarities of March 5th and GRBs prove that GRBs are galactic?* If March 5th is a GRB, then it is the long-sought counterpart (N49) and at least some GRBs would be galactic. This proposition cannot be completely accepted since March 5th and cosmological GRBs might have measured characteristics that overlap, yet be fundamentally different. For example, solar events also sometimes look like classic GRBs, but no one suggests that they are the same physical process.

There seem to be three different physical processes: SGR behavior, March 5th behavior, and GRB behavior. The SGR behavior is characterized by a soft spectrum with typical photon energies of $\sim 30$ keV and fluxes that are $\sim 10^4$ times Eddington. We have seen this behavior in the pulsations following the March 5th initial peak, in the March 5th recurrences, and in SGR 1806-20 and SGR 1900+14. The March 5th behavior is characterized by a somewhat soft GRB spectrum with a possible cutoff below 100 keV and fluxes that are $\sim 10^6$ Eddington. The GRB behavior is characterized by a hard spectrum that extends to GeV energies and fluxes the order of $10^4 - 10^5$ times Eddington (if they are in the galactic corona). It is clear that a single object (a neutron star in N49) gave rise to both SGR and March 5th behavior. One hypothesis is that SGRs escape to the extended galactic corona and become GRBs, meaning that all three processes can occur on the same object. However, we are *not* suggesting that SGRs and GRBs are the same. These phenomena are quite different (by a factor of 10 in the average energy, a factor of $10^3$ in luminosity, a factor of $10^2$ in their duration distributions, and very different repetition rates). We are suggesting that March 5th and GRBs could be the same process and that the SGR and GRB phenomena can occur on the same object but probably the results of different physics.

*5.3 Relationship to Galactic Corona Models*

The March 5th characteristics do not fit into all galactic corona models for GRBs. The central problem with galactic corona models is how to achieve the apparent uniform density of nearby sources as implied by the $-3/2$ portion of the log $N$-log $P$ distribution (Fenimore et al. 1993). Two methods have been suggested: a delayed turn-on of the GRB phase (Li & Dermer 1992) or beaming correlated with the velocity vector (Duncan, Li, & Thompson 1993). In the delayed turn-on models, the high-velocity stars require $\sim 10^{7-8}$ years to reach the outer galactic corona. It is not clear what evolving characteristic changes during the $10^8$ years such that the neutron stars go into a GRB phase. The reputed cyclotron absorption features near 20 keV (Murakami et al. 1988, Fenimore et al. 1988) are often used to argue that GRBs occur on neutron stars with $10^{12}$ G fields (see Lamb 1995). Perhaps the GRB phase requires slow rotators with $10^{12}$ G fields, and to be a high-velocity neutron star requires a $10^{14}$ G field. The $10^8$ yr could be the magnetic field decay time scale and the time it takes to get to the galactic corona. Using March 5th to support the delayed turn-on scenario seems to have two problems, both associated with the young age of N49. First, March 5th would still have its initial high magnetic field since it has had insufficient time to decay. Thus, it does not seem that GRB behavior *must* wait until the field has decayed away, so it is unclear what governs the turn-on time. (March 5th is one of the best cases for a $10^{14}$ G field since that is what is required to slow down the rotation to 8 s [Duncan & Thompson 1992]). Second, since N49 is only $\sim 10^4$ yr old, one would have to propose that March 5th occurred extraordinarily early in the life of the high-velocity neutron star relative to that expected for models with a delayed turn-on. That seems to defeat the purpose of using the March 5th event as an example of a neutron star consistent with the galactic corona model.

The other possible explanation for the apparent uniform density is that the neutron stars are flowing out such that the density of neutron stars is falling off as $r^{-2}$. However, beaming (if correlated with the velocity vector) cancels the $r^{-2}$, giving the appearance of a uniform distribution (Duncan, Li, & Thompson 1993). In this case, one does not have to worry that March 5th occurred too early in the life of the neutron star, but one has to assume that we are, by chance, within a narrow cone (perhaps 20°) in which March 5th radiates. Although the probability that we are within the beam pattern of any one neutron star is small, the expected number could be large and, apparently, March 5th is one of them. The pulse-interpulse structure following the initial peak of March 5th implies that the beaming is not aligned with the rotational axis (although other scenarios are possible). If the beaming is aligned with the velocity vector, then the actual velocity is approximately the projected velocity divided by the sine of the beam angle. The high resulting velocity ($\sim 3500$ km s$^{-1}$), is an argument that the beaming (and, presumably, the magnetic field) is not aligned with the velocity vector.

The mechanism providing the velocity kick could be anisotropic neutrino emission (Duncan & Thompson 1992) The initial velocity vector should be aligned with the spin axis (rather than with the magnetic field) because the spin period is shorter than the duration of the neutrino emission responsible for the high-velocity kick such that the time-averaged kick is along the spin axis (Duncan & Thompson 1992). Thus, the magnetic fields could be initially misaligned from the spin axis. On the time scale of $10^5$ years, the dragging of the neutron star magnetosphere produces a torque which could align the magnetic field with the spin axis and therefore the velocity vector (Michel & Goldwire 1970). Thus, early in the life of the high-velocity neutron star, the magnetic field (and beaming) is not aligned with the velocity. During this phase, the sources would appear to follow the distribution of SNRs. We identify this phase with SGRs. Once the magnetic field is aligned with the spin axis (and velocity vector), the beaming cancels the $r^{-2}$ dependency of the density of high velocity neutron stars and the distribution appears isotropic. We identify this phase with classic GRBs. This scenario does not require a delayed turn-on so, occasionally, a GRB (such as March 5th) occurs while the magnetic field is not aligned with the spin axis. These events can be early enough in the life of the high-velocity neutron star that they can be associated with the parent SNR producing a source within a small error box. When neutron stars are young, the magnetic field has only partially slowed down the spin such that long-period pulsations (e.g., 8 s) are possible. When the neutron star is older and farther out in the corona, the rotation has slowed down to the point that pulsations are not seen in classic GRBs.

In summary, we have reanalyzed the PVO-ICE observations of March 5th and did not find the soft, SGR-like component that was reported by Mazets et al. 1982. We have extensively explored the parameter space of the ICE response function and the corrections to PVO to find a situation where our observations could be consistent with the soft, SGR-like component. All of our analyses indicate a normal GRB spectrum with no soft component. Whether or not March 5th should be considered a classic GRB or a unique event can be argued either way. In either case, the March 5th event demonstrates that neutron stars at distances of tens of kpc are capable of producing the energy release, time history, and spectrum that is required of GRBs in an extended galactic corona.

*Acknowledgements:* The authors thank D. Lamb, B. Paczyński, R. Epstein, H. Li, and C. Kouveliotou for useful comments. This work was done under the auspices of the US Department of Energy and was supported in

# FIGURE CAPTIONS

**Fig. 1:** A comparison of GRB spectra, an SGR spectrum, and the Konus March 5th spectrum. The solid curves are 54 GRB spectra observed by BATSE (Band et al. 1993). The histogram is the Konus spectrum taken from Mazets et al. 1982. The dotted curve is the SGR 1806-20 spectrum (Fenimore, Laros, & Ulmer 1994). The Konus spectrum has a soft component similar to that of SGRs.

**Fig. 2:** Fits of the Konus spectrum for March 5th to the ICE observations. The upper panel is the Konus spectrum reported by Mazets et al. 1982. The bottom panel gives the residuals between the ICE observations and the Konus spectrum folded through a range of ICE response matrices. The residuals are in units of the uncertainty of the observations (the $1\sigma$ error bars on the points in the bottom panel are equal to $\pm 1$ by definition). The large residues with clear trends indicate that the Konus spectrum is inconsistent with the ICE observations.

**Fig. 3:** The range of possible spectra that can fit the ICE observations assuming 1 cm of aluminum in the central core of ICE. The solid line in the upper panel is the best-fit, three-segment power law spectrum, and the solid squares in the bottom panel give the corresponding residues. The small residues with no clear trend indicate an acceptable fit. The dotted, dashed, and long dashed curves (and open squares) represent spectra that fit the data at $1\sigma, 2\sigma$, and $3\sigma$, respectively. A soft component such as that reported from the Konus experiment is not consistent with the ICE observations at least to the $3\sigma$ level.

**Fig. 4:** The possible spectra that can fit the ICE observations assuming a range of materials in the satellite. From bottom to top, the curves are for 1/4, 1/2, 1, 2, and 4 cm of aluminum in the central core of the ICE satellite. This range of aluminum spans all reasonable response matrices. The bottom panel gives the corresponding residues. All fits are consistent with the observations, but none are consistent with the Konus soft component.

**Fig. 5:** Fits to the PVO March 5th observations. The dotted curve in the upper panel and the open squares in the bottom panel (residues) assume a power law spectrum. The solid curve and solid squares are for a power law with a high energy cutoff. The addition of a single parameter (the cutoff

energy) reduces a goodness-of-fit statistic from 155 for the power law to 10 for the power law with cutoff. The long dashed curve and open circles are for a thermal bremsstrahlung shape; the temperature is 246 keV; and the $\chi^2$ is 6.5. The dashed line and open triangles are for two connected power laws; the $\chi^2$ is zero because there are no degrees of freedom.

**Fig. 6:** Fit of the ICE and PVO observations of March 5th to a power law connected to a thermal bremsstrahlung-like spectral shape. The best-fit function is the solid line in the top panel, and the solid and open points in the bottom panel are the residuals for ICE and PVO, respectively. The dotted, dashed, and long-dashed curves in the top panel represent spectra that fit the data at the $1\sigma, 2\sigma$, and $3\sigma$ confidence levels, respectively.

TABLE 1
Intensity and Luminosity of the March 5th event on Various Time Scales

|  | 30 - 2000 keV | | 50 - 300 keV | | 100 - 500 keV | |
| --- | --- | --- | --- | --- | --- | --- |
|  | $P$ | $L_{45}$ | $P$ | $L_{45}$ | $P$ | $L_{45}$ |
| PVO[1] 10 ms | $2.1 \times 10^4$ | 1.9 | $1.2 \times 10^4$ | 0.87 | $7.9 \times 10^3$ | 1.0 |
| PVO[1] 64 ms | $1.8 \times 10^4$ | 1.4 | $1.0 \times 10^4$ | 0.71 | $6.3 \times 10^3$ | 0.80 |
| ICE-PVO[2] 256 ms | $8.9 \times 10^3$ | 0.61 | $5.0 \times 10^3$ | 0.32 | $3.0 \times 10^3$ | 0.35 |
| Konus[3] 200 ms | $1.1 \times 10^4$ | 0.51 | $4.7 \times 10^3$ | 0.24 | $1.6 \times 10^3$ | 0.18 |

[1]cf. equation (4)
[2]cf. equation (3)
[3]cf. equation (2)

Figure 1

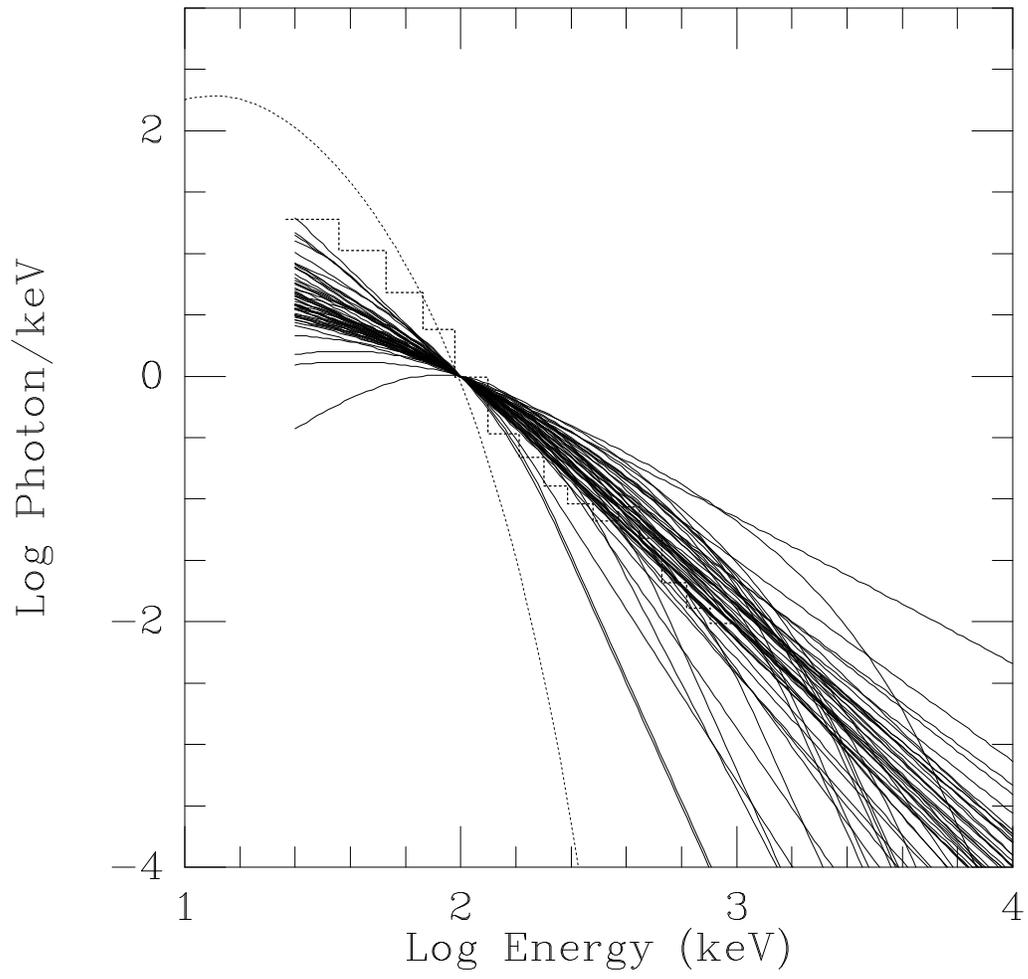

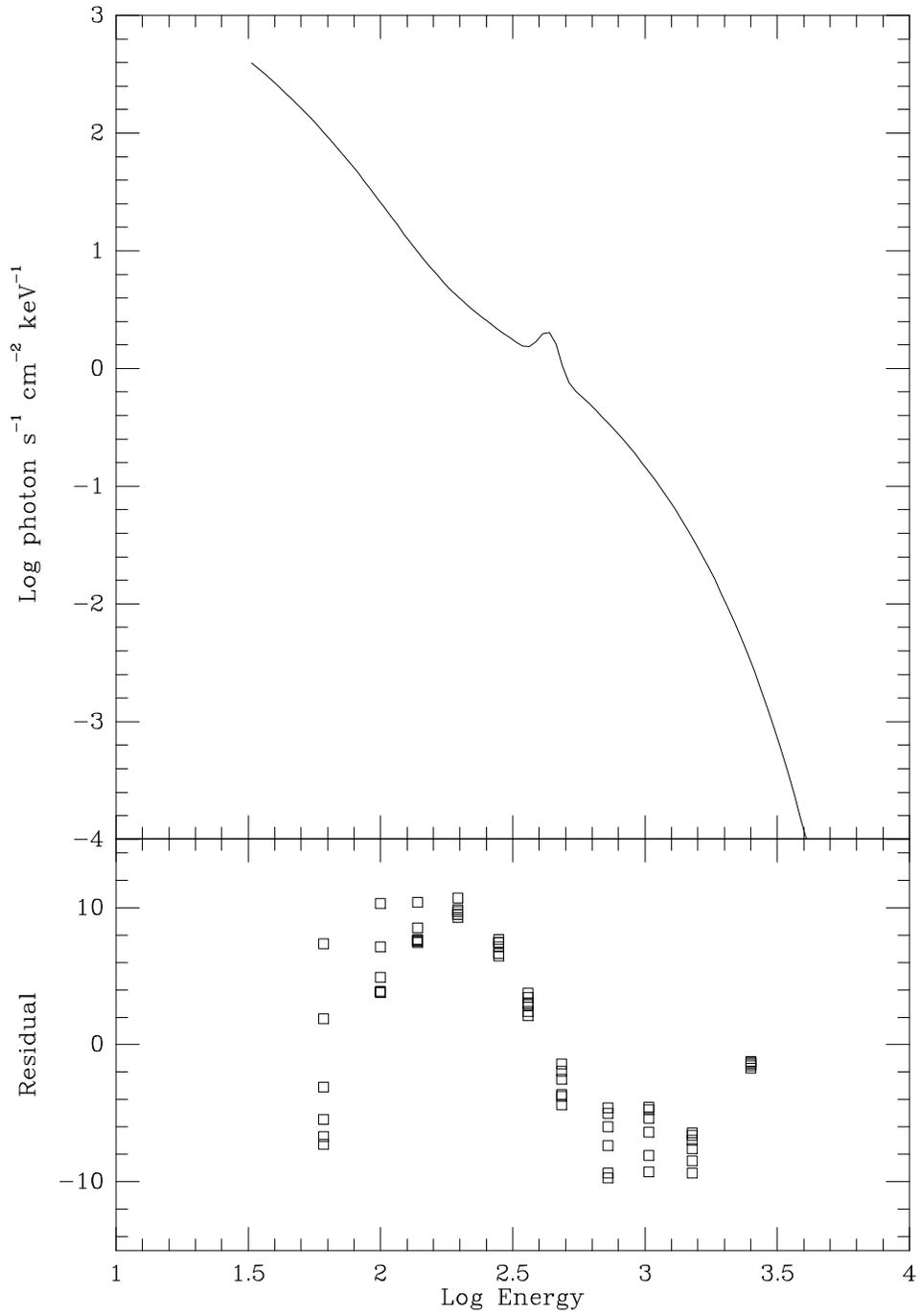

Figure 2

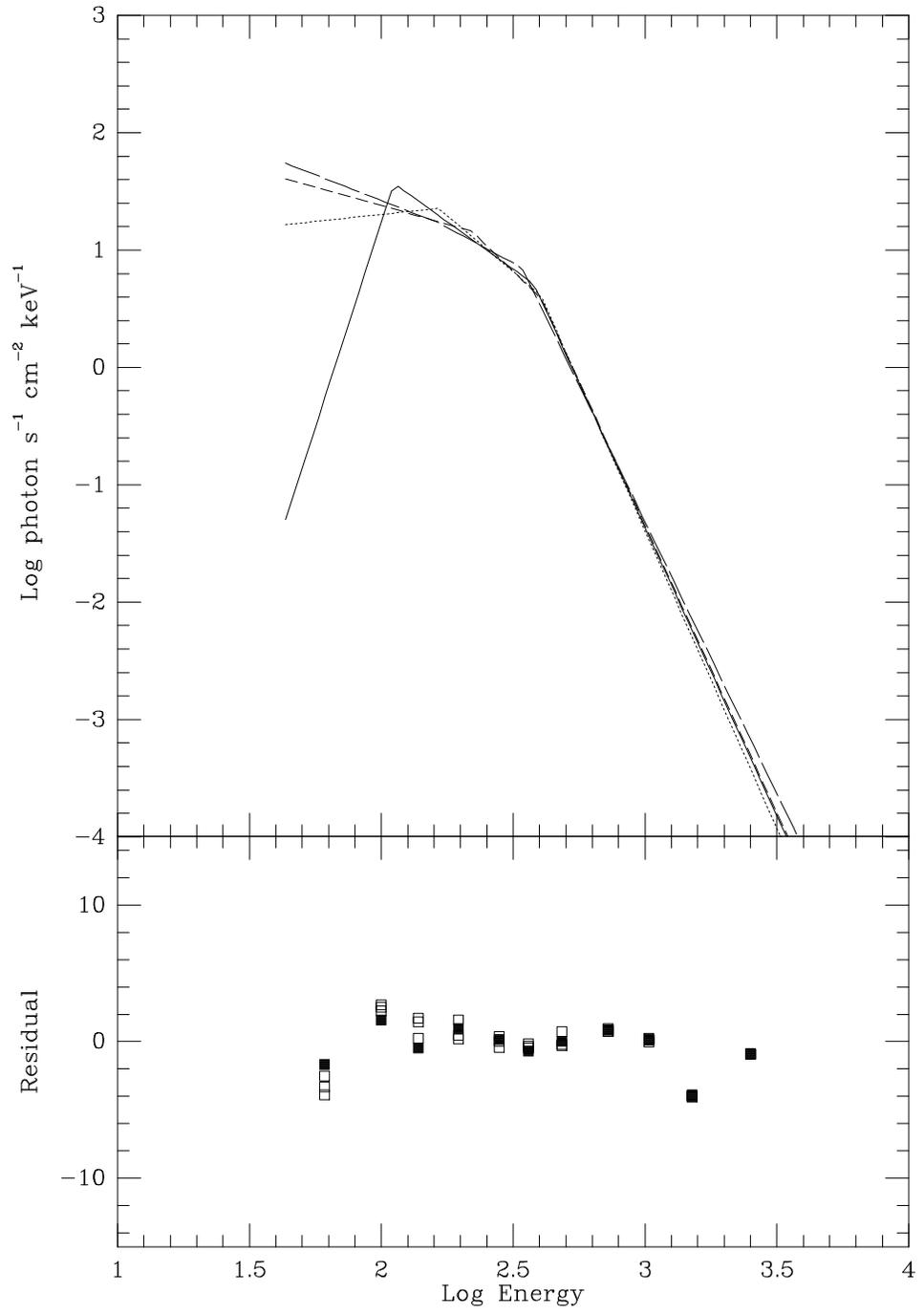
Fig 3

Figure 4

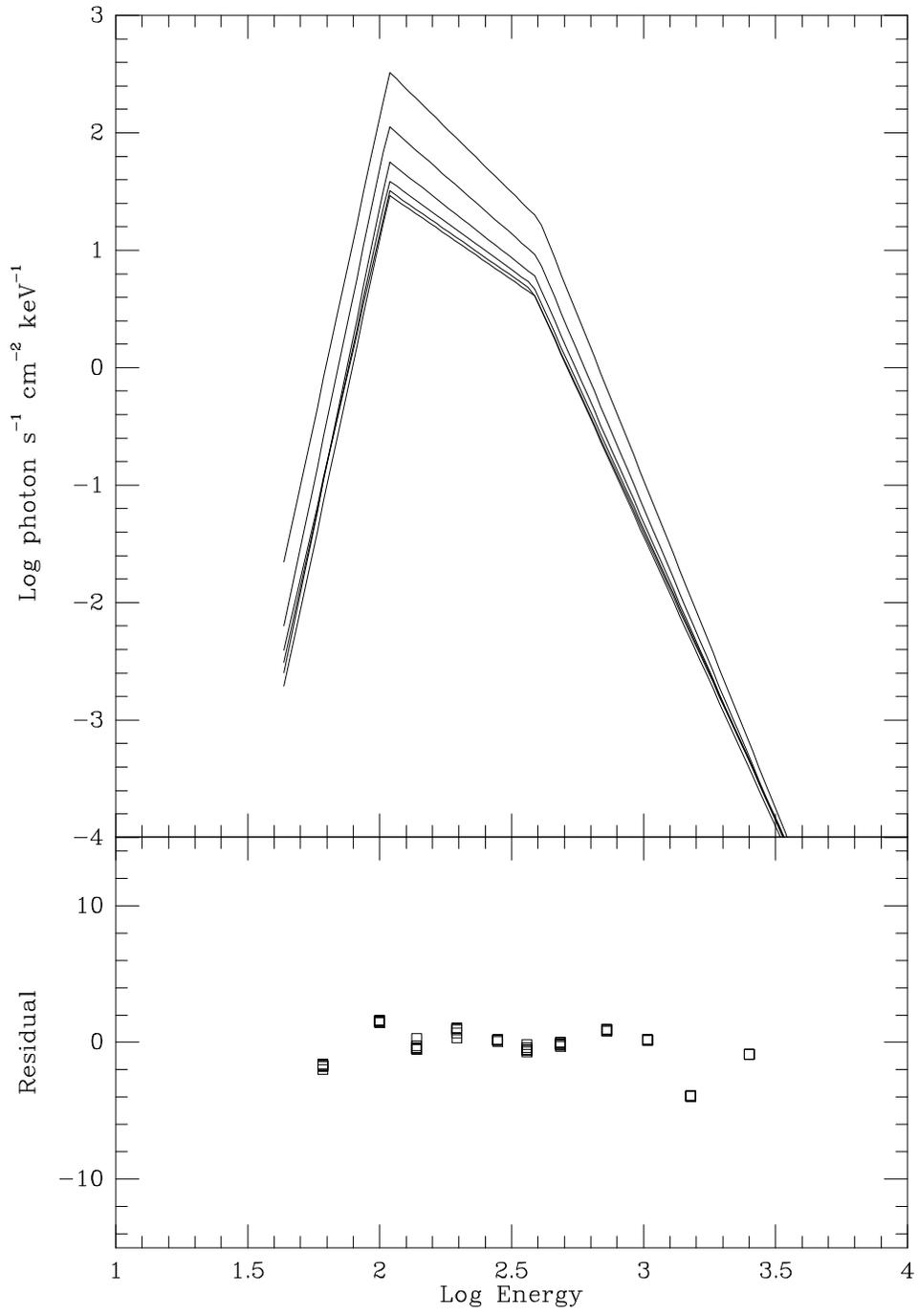

Figure 5

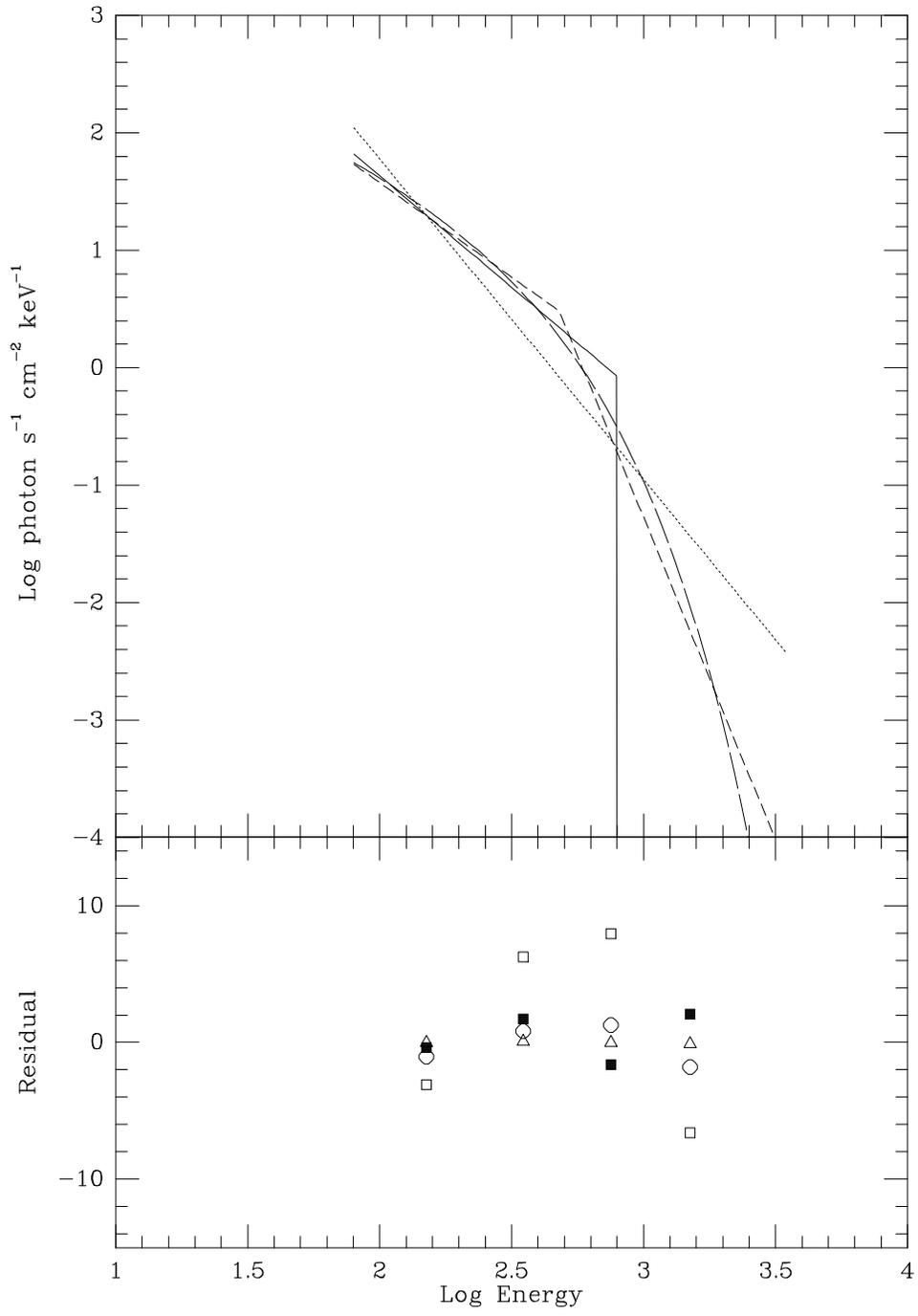

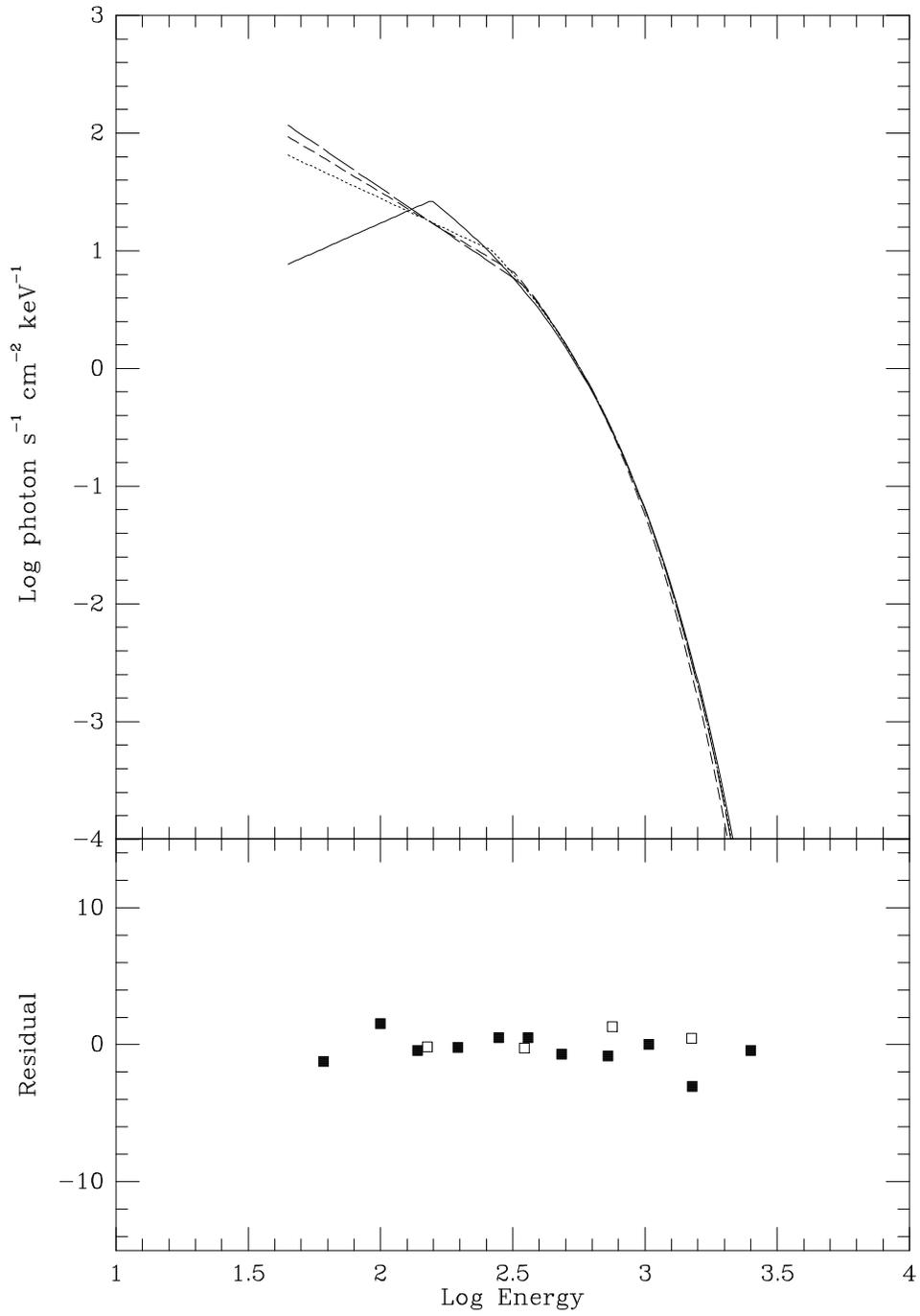

Figure 6